\documentclass[%
 reprint,
 amsmath,amssymb,
 aps,
]{revtex4-2}

\usepackage{graphicx}% Include figure files
\usepackage{dcolumn}% Align table columns on decimal point
\usepackage{bm}
\usepackage{subfig}
\usepackage{multirow}
\usepackage[utf8]{inputenc}
\begin{document}

\preprint{APS/123-QED}

\title{Swarm of lightsail nanosatellites for Solar System exploration} 

\author{Giovanni Santi}
\affiliation{Università di Padova, Centro di Ateneo di Studi e Attività Spaziali (CISAS), via Venezia, 15, 35131 Padova, Italy}
\author{Alain Jody Corso}
\affiliation{Consiglio Nazionale delle Ricerche - Istituto di Fotonica e Nanotecnologie (CNR-IFN), via Trasea, 7, 35131 Padova, Italy}
\author{Denis Garoli}
\affiliation{Istituto Italiano di Tecnologia, Via Morego 30, 16163 Genova, Italy}
\author{Giuseppe Emanuele Lio}
\affiliation{Dipartimento di Fisica e European Laboratory for Non-Linear Spectroscopy (LENS), Università di Firenze, Via Nello Carrara 1, 50019, Sesto Fiorentino, Florence, Italy}
\author{Marco Manente}
\affiliation{Technology for Propulsion and Innovation S.p.A. (T4i), Via Emilia, 15, 35043 Monselice, Padova, Italy}
\author{Giulio Favaro}
\affiliation{Università di Padova, Dipartimento di Fisica e Astronomia, via Marzolo 8, 35131 Padova, Italy } 
\author{Marco Bazzan}
\affiliation{Università di Padova, Dipartimento di Fisica e Astronomia, via Marzolo 8, 35131 Padova, Italy}
\author{Giampaolo Piotto}
\affiliation{Università di Padova, Dipartimento di Fisica e Astronomia, via Marzolo 8, 35131 Padova, Italy\\Università di Padova, Centro di Ateneo di Studi e Attività Spaziali (CISAS), via Venezia, 15, 35131 Padova, Italy}
\author{Nicola Andriolli}
\affiliation{Consiglio Nazionale delle Ricerche, Istituto di Elettronica, Ingegneria dell'Informazione e delle Telecomunicazioni, via Caruso 16, 56122 Pisa, Italy }
\author{Lucanos Strambini}
\affiliation{Consiglio Nazionale delle Ricerche, Istituto di Elettronica, Ingegneria dell'Informazione e delle Telecomunicazioni, via Caruso 16, 56122 Pisa, Italy}
\author{Daniele Pavarin}
\affiliation{Università di Padova, Dipartimento di Ingegneria Industriale, via Gradenigo, 6A, 35131 Padova, Italy \\ Centro di Ateneo di Studi e Attività Spaziali (CISAS), via Venezia, 15, 35131 Padova, Italy}
\author{Leonardo Badia}
\affiliation{Università di Padova, Dipartimento di Ingegneria dell'Informazione, via Gradenigo, 6B, 35131 Padova, Italy}
\author{Remo Proietti Zaccaria}
\affiliation{Istituto Italiano di Tecnologia, Via Morego 30, 16163 Genova, Italy }
\author{Philip Lubin}
\affiliation{Department of Physics, University of California - Santa Barbara, CA, 93106}
\author{Roberto Ragazzoni}
\affiliation{Università di Padova, Dipartimento di Fisica e Astronomia, via Marzolo 8, 35131 Padova, Italy \\ Istituto Nazionale di Astrofisica, Osservatorio Astronomico di Padova, Vicolo dell'Osservatorio, 5, 35122 Padova, Italy}
\author{Maria G. Pelizzo}
\email{mariaguglielmina.pelizzo@unipd.it}
\affiliation{Università di Padova, Dipartimento di Ingegneria dell'Informazione, via Gradenigo, 6B, 35131 Padova, Italy \\ Centro di Ateneo di Studi e Attività Spaziali (CISAS), via Venezia, 15, 35131 Padova, Italy}

\date{\today}% It is always \today, today,

\begin{abstract}
This paper presents a study for the realization of a space mission which employs nanosatellites driven by an external laser source impinging on an optimized lightsail, as a valuable technology to launch swarms of spacecrafts into the Solar System. Nanosatellites propelled by laser can be useful for the heliosphere exploration and for planetary observation, if suitably equipped with sensors, or be adopted for the establishment of network systems when placed into specific orbits. By varying the \textit{area-to-mass ratio} (i.e., the ratio between the sail area and the payload weight) and the laser power, it is possible to insert nanosatellites into different hyperbolic orbits with respect to Earth, thus reaching the target by means of controlled trajectories in a relatively short amount of time. A mission involving nanosatellites of the order of 1 kg of mass is envisioned, by describing all the on-board subsystems and satisfying all the requirements in term of power and mass budget. Particular attention is paid to the telecommunication subsystem, which must offer all the necessary functionalities. To fabricate the lightsail, the thin films technology has been considered, by verifying the sail thermal stability during the thrust phase. Moreover, the problem of mechanical stability of the lightsail has been tackled, showing that the distance between the ligthsail structure and the payload plays a pivotal role. Some potential applications of the proposed technology are discussed, such as the mapping of the heliospheric environment.
\end{abstract}

\maketitle

%\tableofcontents

\section{Introduction}

\begin{figure*}
    \centering
    \includegraphics[width=0.7\textwidth]{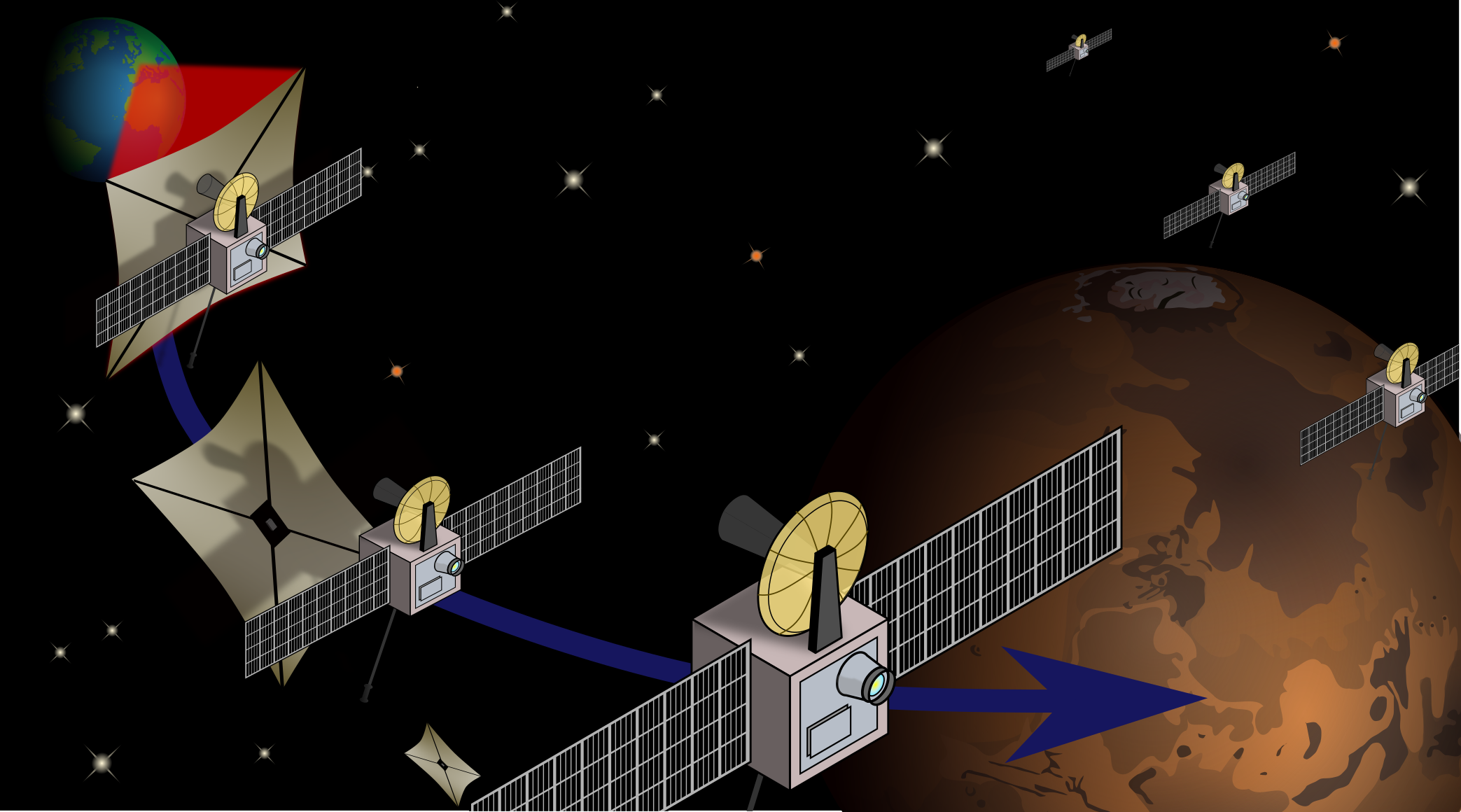}
    \caption{Concept of swarm of nanosatellite propelled by laser.}
    \label{fig:Graphics}
\end{figure*}

The increase of human activities in space, accompanied by the desire of interplanetary travels for scientific and commercial purposes, requires access to small, inexpensive, and easy-to-launch satellites. In order to keep the costs as low as possible, nanosatellite configurations should be the choice of reference, as they are synonymous of low mass. However, the most common propulsion technologies rely on large amount of chemical propellant, which needs to be carried by the spacecraft, thus increasing the mass and launch costs. For these reasons great attention has been raising towards the concept of propellant-free satellites, mainly referring to the use of solar sails. The idea of using sails to move objects in space became the subject of research in the mid-1950s, when the first spacecraft propelled by the pressure of solar radiation was conceived. Since then, many theoretical studies have been focusing on the key technologies characterizing the solar sail propulsion: structure and materials forming the sail, the deployment phase system, orbital dynamics and attitude control systems \cite{Gong2019, Sprowitz2019}. However, very few lab-scale experimental tests have been carried out, with an even lower amount of missions actually launched into space \cite{Whorton2008, PlanetarySoc2019, Mori2010}. These missions have been especially focusing on the deployment phase of the sail and the operation of the related subsystems, rather than on the use of the sail as a propulsion system. Indeed, some problems remain due to the limited power of the Sun and the directionality of the thrust, the latter requiring sophisticated attitude control solutions. These deficiencies could be overcome by laser-based propulsion systems, eventually realized as ultra-high-power laser arrays, which offer thrusts of remarkable magnitude, and the ability to act on the direction of the thrust vector. Herein we shall refer to laser-driven solar sails as \textit{lightsail} and to the thrusting lasing technology as \textit{Direct Energy Laser Propulsion} (${\mathrm{DELP}}$). More in general, the laser-based technology can either be propellant-based \cite{Levchenko2018} or propellant-free. In the first case  the thrust is due to the impulse provided by a flow of particles ejected by ablation \cite{Felicetti2013, Zhang2015, Phipps2010}, whereas in the second case the thrust is provided by the momentum exchange with the incident photons \cite{Marx1966,Forward1984,Lubin2016,Kulkarni2018,Lubin2020}. The main advantage the propellant-free approach is that the energy source is completely separated from the spacecraft itself, with the strong advantage of not requiring any propellant onboard. 
Recently, ${\mathrm{DELP}}$ has been proposed as a way to achieve relativistic velocities necessary to cover deep-space distances in a short period of time, as foreseen by the ${\mathrm{NASA}}$ \textit{Starlight} \cite{Starlight2021} program and the \textit{Breakthrough Initiative} \cite{Brekthrough2021}. The goal of these projects is demonstrating the feasibility of the first interstellar missions, by targeting other stellar systems such as Alpha Centauri \cite{Lubin2016,Perakis2016, Kulkarni2018,Lubin2020,Kudyshev2021}. A laser system placed on the ground has been conceived as a realistic launch base which can recurrently send spacecrafts into space, making its implementation cost-effective \cite{Hilase2021}. The use of laser arrays enables modularity and scalability, both necessary ingredients for the achievement of extremely high power, which in some studies is expected to be in the gigawatt range \cite{Lubin2016, Parkin2018, Lubin2020,Worden2021}. In this respect, studies on the phase control for the coherent combination of beams are in progress \cite{Bandutunga2021}. The use of a high-power laser as a propulsive source brings into play the radiation pressure exerted on the lightsail, with values of few orders of magnitude greater than what envisaged for solar sails. This important difference determines obvious mechanical and thermal effects, which require dedicated studies on sail stability and sail composition \cite{Davoyan2021, Campbell2022, Santi2022, Rafat2022, Savu2022}. In particular, thin film and multilayer sails, photonic crystals, gratings and metasurfaces have been recently introduced to improve the efficiency and the stability of lighsails \cite{Manchester2017, Swartzlander2017,  Ilic2018, Atwater2018, Chu2019, Ilic2019, Srivastava2019, Siegel2019, Salary2020, Shirin2021, Holdman2021, Gieseler2021, Salary2021, Kumar2021, Gao2022, Brewer2022}. 
Furthermore, it is important to quantify the damage that the interstellar medium can induce on the lightsail and the payload \cite{Drobny2021}.
Although many proposed mission scenarios (e.g., ${\mathrm{NASA}}$ \emph{Starlight}) foresee the use of low-mass solutions, it is still unclear how these will be able to host a telecommunication system fully capable of data communication and possible telemetry from deep space. In fact, the need to communicate over large distances requires the use of adequate antennas in terms of power and gain. In order to limit the power demand as much as possible, the use of very narrow lobes could be an option, with however the drawback of requiring an attitude control system. 

In the present work all these aspects are addressed and discussed with the aim of enabling a new ${\mathrm{DELP}}$-based mission, conceived to launch swarms of small satellites at non-relativistic speeds ($v \ll c$) to travel the Solar System, hence exploring the heliosphere and targeting planets \ref{fig:Graphics}. ${\mathrm{DELP}}$ will be assumed as the baseline technology, because its development and use for Solar System missions is a fundamental prerequisite to enable deep space missions. In this regard, the power and telecommunication requirements for Solar System missions are less demanding than for deep space missions, even though many technological aspects remain challenging. Furthermore, by considering different mission targets, advantages and limits of the proposed technology will be discussed with respect to another proposed technology concept, the \textit{Laser Electric Propulsion} (${\mathrm{LEP}}$). Here a laser is employed to supply energy to a photovoltaic system, which in turn generates the necessary electricity to power an electric propulsion system~\cite{Sheerin2021}. Differently from ${\mathrm{DELP}}$, ${\mathrm{LEP}}$ requires each nanosatellite equipped with its own propulsion system, with increasing cost when a large number of satellites are launched. The idea of exploring Mars by the use of swarm of nanosatellites, has been recently conceived \cite{Timmons2021, Duplay2022, Sheerin2021}, including the case of propellant-based laser propulsion technology \cite{Duplay2022}.  Vice versa, in the present work, we propose the first planetary mission based on nanosatellites propelled by ${\mathrm{DELP}}$ by carefully analysing all related technical aspects. Using ${\mathrm{DELP}}$ technology, payloads of the order of 1 kg of mass can reach a planet like Mars or Venus in few days. The non-relativistic speed reduces the demand in terms of laser power, making the mission more feasible. The weight of the payload is also pivotal to mechanically stabilize the complete nanosatellite system (lightsail plus payload), thus removing all the technological challenges related to stability posed by a few grams satellite for interstellar travel \cite{Popova2017}. Besides careful considerations on the power system, the technology for the realization of the lightsail is also discussed, assuming the thin film solution as the baseline \cite{Ilic2018, Santi2022}. In this respect, the mechanical and thermal stability of the lightsail during the thrust phase are analyzed. Finally, potential scientific applications of the mission are presented. 

\section{Propulsion and Dynamics}
\begin{figure}
    \centering
     \includegraphics[width=.5\textwidth]{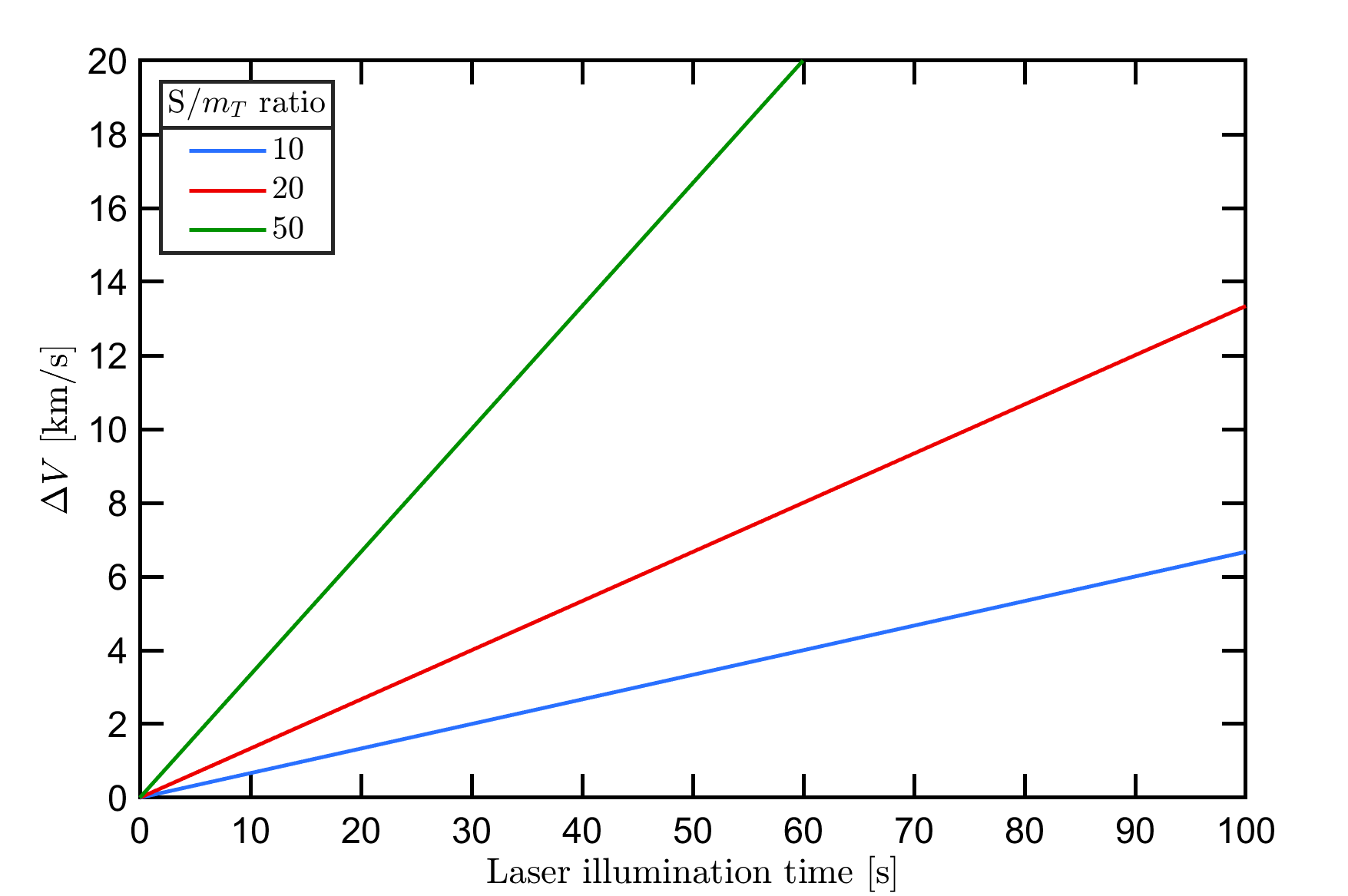}
    \caption{Dependence of the velocity gain $\Delta V$ from the laser illumination time for different $S/m_\mathrm{T}$ values. The simulation assumes $I_0=1$ GW m$^{-2}$ and departure from a geostationary orbit. }
    \label{fig:DV}
\end{figure}

\begin{figure}
    \centering
     \includegraphics[width=.5\textwidth]{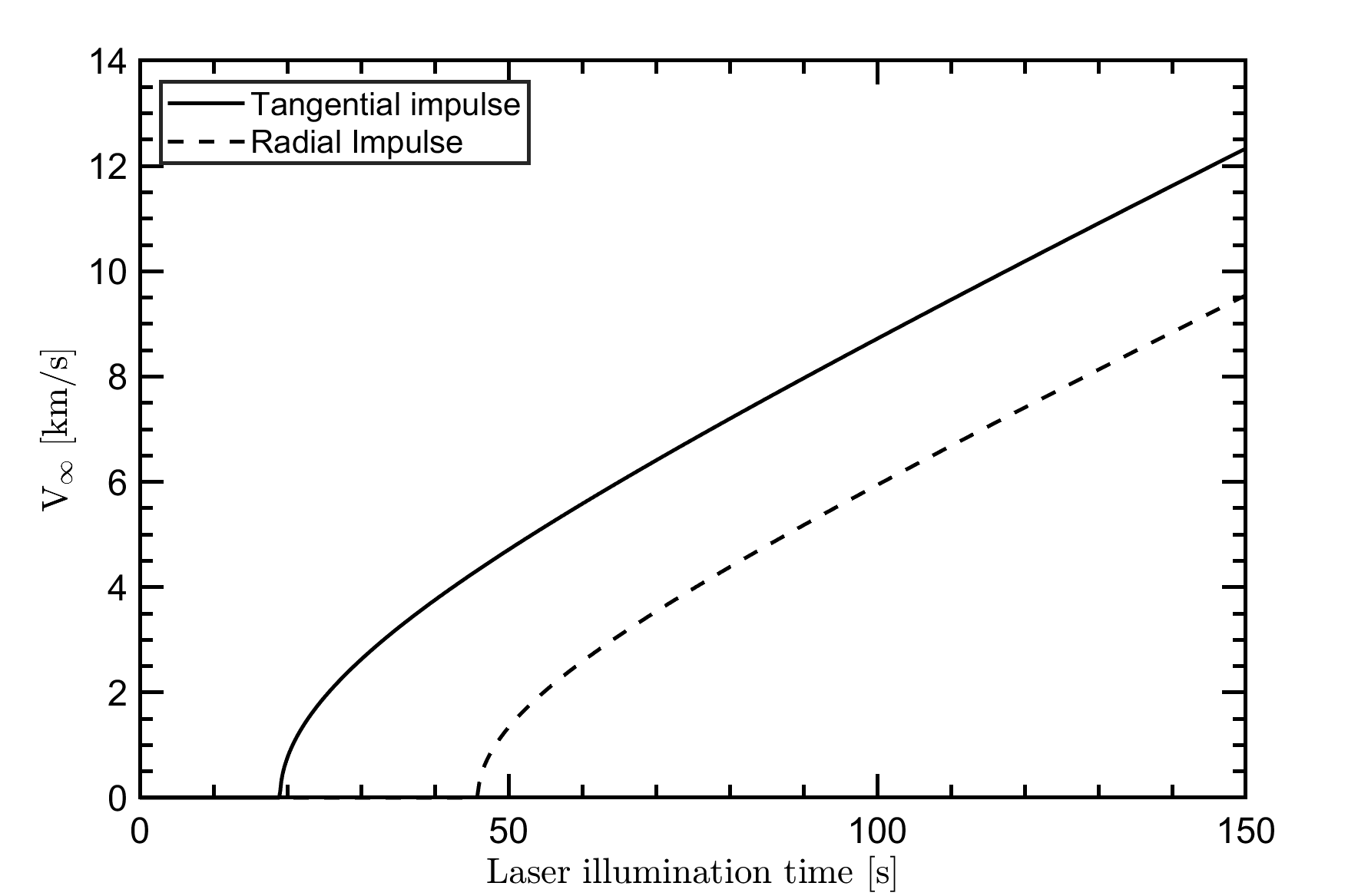}
    \caption{Hyperbolic excess velocity for a tangential and a radial impulse. The simulation assumes an area to mass ratio ($S/m_\mathrm{T}$) equal to 10, $I_0=1$ GW m$^{-2}$ and departure from a geostationary orbit.}
    \label{fig:VInf}
\end{figure}

The key principle of propulsion in space is the need of expelling part of the mass carried by the spacecraft in order to produce thrust. Indeed, according to the Newtonian law of motion, the force \textit{F} generated by a propulsion system has magnitude $\dot{m}V_\mathrm{ext}$, being $\dot{m}$ and $V_\mathrm{ext}$ the mass consumption rate and the exhaust velocity respectively, and has opposite direction in respect to the ejected particles. Space propulsion systems are then typically classified according to the physical processes used to eject the mass: chemical systems exploit a chemical reaction between solid and/or liquid propellants, while electric/magnetic systems accelerate ions through a tailored electric/magnetic field that mimics the behaviour of a nozzle. 
Nevertheless, this concept describes a \textit{classical} propulsion system. Indeed, when photons are the exhausted particles, the concept of mass consumption rate does not hold any longer. In a laser-driven propulsion system, the force exerted on a flat non-diffusing lightsail can be written as \cite{Atwater2018}:

\begin{equation}\label{F_L}
    F_\mathrm{L} = \frac{I_0S}{c}\left[2R(\theta_i,\lambda_0)\cos{\theta_i}\hat{n} + A(\theta_i,\lambda_0)\hat{d}\right]
\end{equation}

where $I_0$ is the laser irradiance, $S$ is the lightsail surface, $R$ and $A$ are the reflectance and absorbance of the lightsail computed at the angle of incidence $\theta_i$ and laser wavelength $\lambda_0$, $\hat{n}$ is the normal to the surface, $\hat{d}$ is the radial unit vector along the laser direction, and $c$ is the speed of light. Eq. \eqref{F_L} provides the general expression, though for practical applications high reflectivity is required to maximize the force, so that the absorption contribution can be neglected. Note also that, while in the case of the relativistic lightsails the Doppler-shift wavelength of the laser source has to be taken into account~\cite{Atwater2018}, in the case of a lightsail used to drive an object at velocities $v \ll c$ the spectral performance can be optimized at the sole laser wavelength. \\
In the following discussion, it is assumed that a nanosatellite is accelerated from a circular parking orbit around Earth. In order to escape the gravitational attraction, the nanosatellite needs to increase its velocity by a \textit{velocity gain} $\Delta V$ and to be inserted into a hyperbolic trajectory arriving at the Earth Sphere of Influence (${\mathrm{SOI}}$) with an \textit{hyperbolic excess velocity} $v_\infty$ greater then zero. Under the assumption of a normal incidence reflection and considering a laser illumination time $t$, the $\Delta V$ can be calculated from Eq. \eqref{F_L}:

\begin{equation}\label{eq:DV}
    \Delta V = \frac{F_\mathrm{L}}{m_\mathrm{T}}t = 2\frac{I_0}{c}\frac{S}{m_\mathrm{T}}t\hat{n}
\end{equation}

where $m_\mathrm{T}$ is the spacecraft mass (i.e., the sum of the lightsail and nanosatellite masses). As shown in Figure \ref{fig:DV}, when the laser is turned on at time $t=0$ s, $\Delta V$ linearly increases with the laser illumination time following a slope depending on $S$, $m_\mathrm{T}$, and $I_0$. Upon fixing the impinging laser irradiance $I_0$, the dependence on the \textit{area-to-mass ratio} ($S/m_\mathrm{T}$) has a direct impact on the performance of the accelerating spacecraft since it determines the time interval that the laser has to be activated to reach the desired $\Delta V$. In turn, this activation time determines the costs associated with the launching phase. 
For instance, by assuming $I_0=1$ GW m$^{-2}$ and departure from a geostationary orbit, the time to achieve a $\Delta V$ of 5 km s$^{-1}$ is $\simeq 75$ s if $S/m_\mathrm{T}=10$ m$^2$ kg$^{-1}$, but only $\simeq 15$ s if $S/m_\mathrm{T}=50$ m$^2$ kg$^{-1}$. In general, a higher $S/m_\mathrm{T}$ value is always to be preferred since it involves a higher thrust efficiency. Moreover, this value is a constraint on the spacecraft design as, for a given irradiance $I_0$, it defines the dimension $S$ of the lightsail necessary to accelerate a mass $m_\mathrm{T}$.
\\
The location of the laser source is presently an open question for the scientific community. Many authors take the conservative assumption of having DELP placed on the Earth surface \cite{Atwater2018, Duplay2022, Ilic2018, Sheerin2021}, an approach typically referred to as ${\mathrm{DELTA}}$ (Directed Energy Launch Technology Array). Nevertheless, an orbiting system named ${\mathrm{DE}}$-${\mathrm{STAR}}$ (Directed Energy System for Targeting of Asteroids and exploRation) has also been proposed \cite{Hughes2014}. This concept mainly derives from the need for a defence system against potential asteroid impacts, but it is intrinsically suitable also for realizing a laser propulsion solution. In the ${\mathrm{DELTA}}$ scenario, the spacecraft experiences a radial ($\bot$) acceleration with respect to the orbit, while in the ${\mathrm{DE}}$-${\mathrm{STAR}}$ approach the spacecraft accelerates tangentially ($\lVert$). Hence, the hyperbolic excess velocity at the Earth ${\mathrm{SOI}}$ can be expressed as: 
\\
\begin{equation}
\begin{split}
    (v^\lVert_\infty)^2 &= \left(v_0 +\Delta V\right)^2 -2\mu\left[ r_0^2+t^2\left(v_0+\frac{\Delta V}{2}\right)^2 \right]^{-1/2} \\
    (v^\bot_\infty)^2 &= v_0^2 +\Delta V^2 -2\mu\left[ \left(r_0+\frac{\Delta Vt}{2}\right)^2+(v_0t)^2  \right]^{-1/2} 
\end{split}
\end{equation}

where $v_0$ and $r_0$ are the velocity and the radius of the circular parking orbit, $t$ is the laser illumination time, $\Delta V$ the velocity gain as defined in Eq. \eqref{eq:DV} and $\mu$ is the standard gravitational parameter of the Earth. As shown in Figure \ref{fig:VInf}, a tangential impulse is generally more efficient than a radial one. Indeed, in the first case the instantaneous velocity of the circular parking orbit and the acceleration direction are parallel to each other, and less energy is required to reach the target $\Delta V$.  
\\

\begin{figure*}
    \centering
\includegraphics[width=1\textwidth]{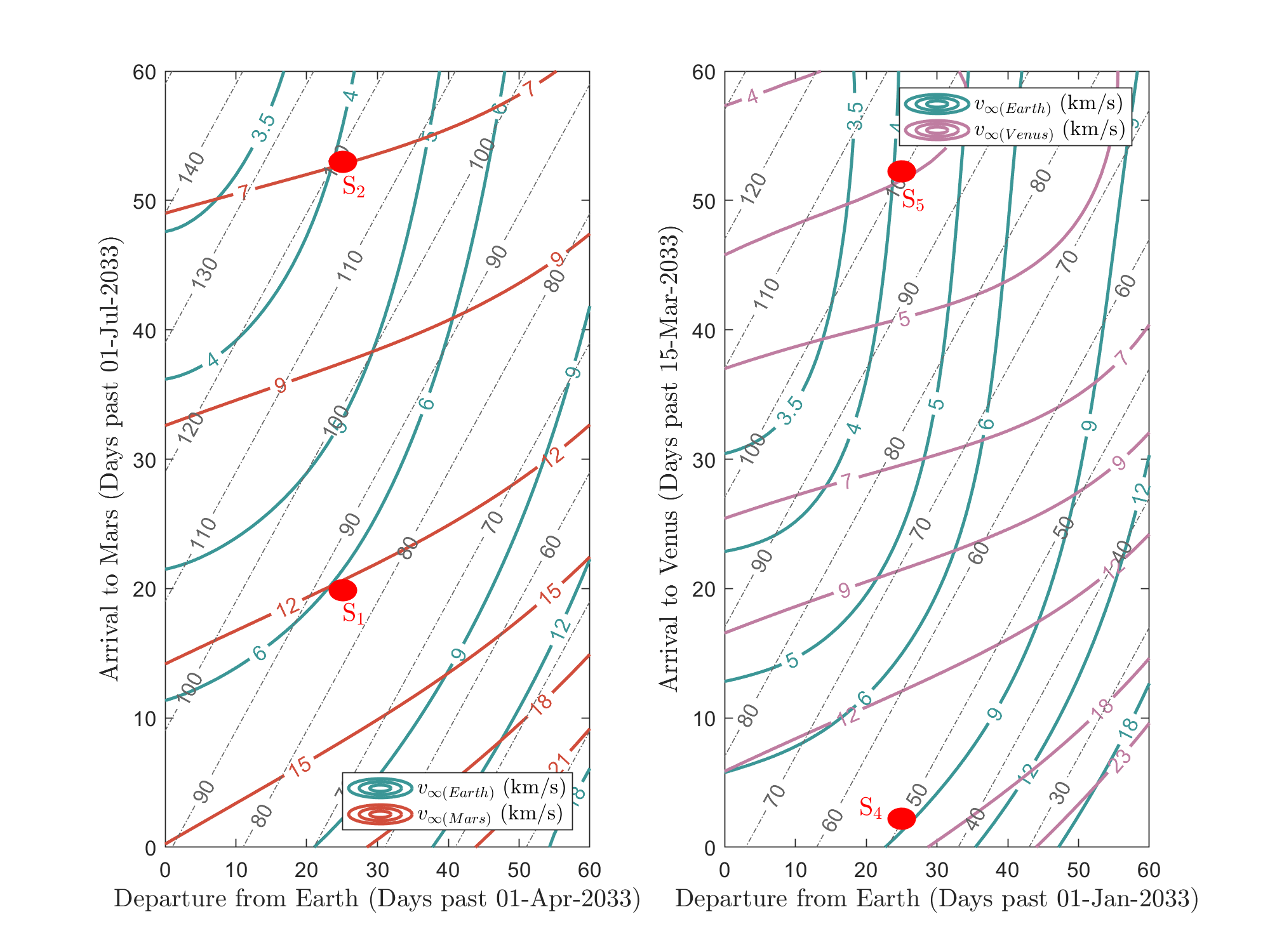}
    \caption{Porkchop plots of hyperbolic excess velocities v$_\infty$ in the case of Earth departure in a 2033 launch window for Mars (left) and Venus (right). The curves of ${\mathrm{TOF}}$ reported are given in step of ten days and highlighted by the dashed grey lines.}
    \label{fig:Porkchop}
\end{figure*}

\begin{figure*}
    \centering
    \subfloat{\includegraphics[width=0.45\textwidth]{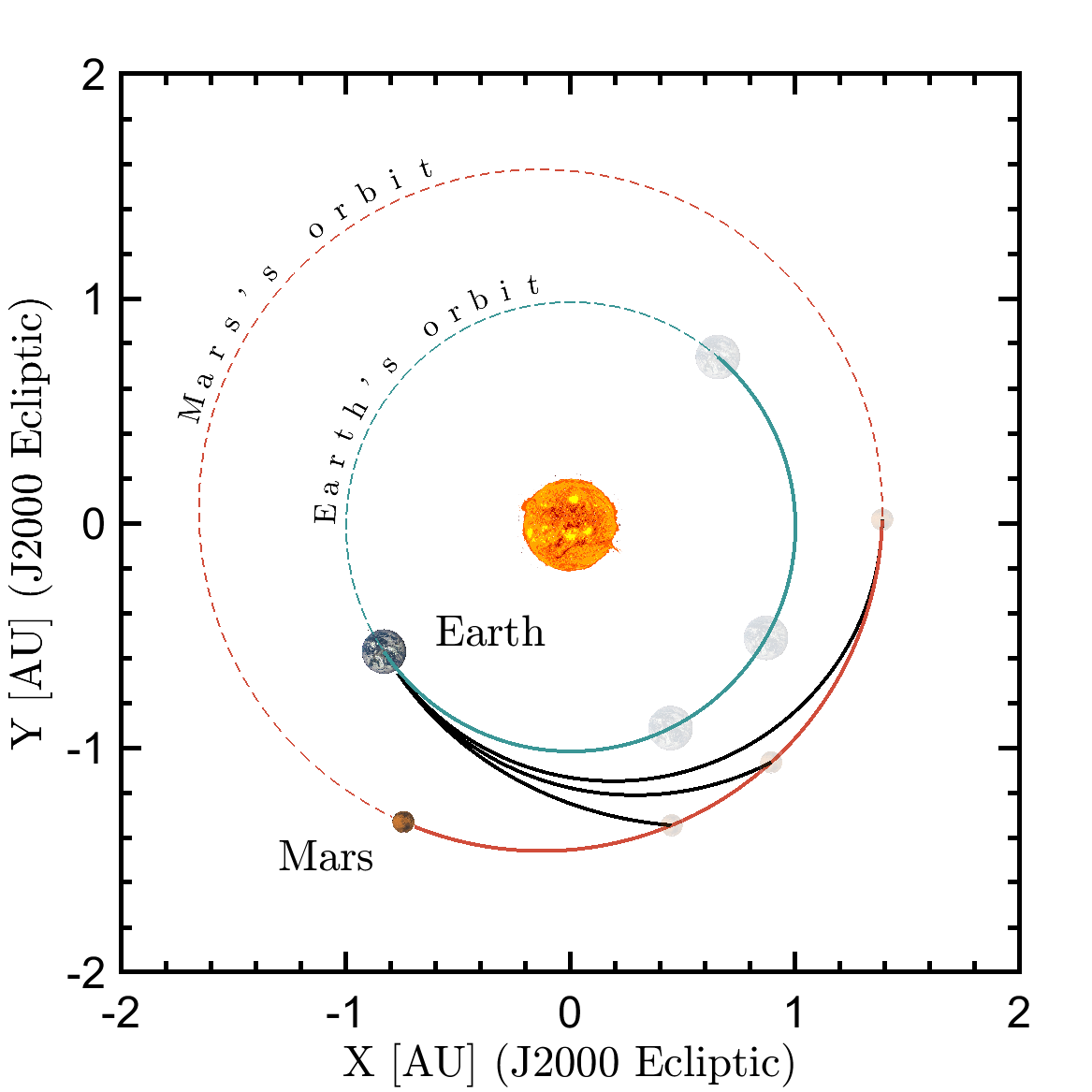}}
    \subfloat{\includegraphics[width=0.45\textwidth]{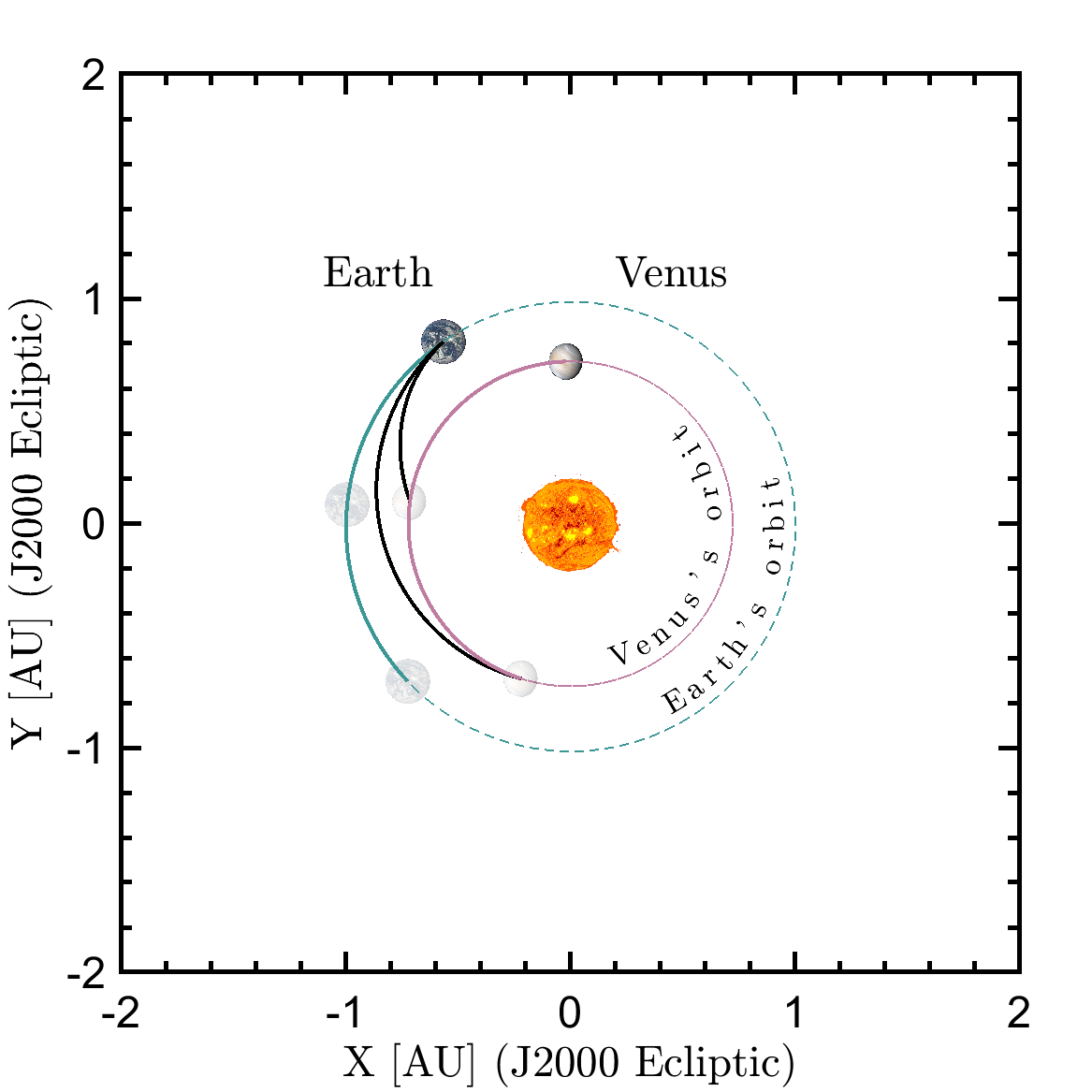}}
    \caption{Orbital transfers to Mars (left) and Venus (right) for the parameters reported in Table \ref{tab:Parameters}. The planets at departure and arrival are shown in full and faded colors, respectively. The portion of the orbits covered by the planet while the spacecraft is travelling into the interplanetary medium are reported in solid lines, and the remaining portion in dashed line.}
    \label{fig:Earth2planet}
\end{figure*}

In order to investigate the breakthrough capability of laser propulsion, the case of a spacecraft with an area to mass ratio $S/m_\mathrm{T}$ of 10 m$^2$ kg$^{-1}$ travelling towards Mars and Venus propelled by a beam of irradiance of $I_0=1$ GW m$^{-2}$ is considered. Assuming a 2033 launch window, the astrodynamics solutions are found by solving the Lambert's problem with a custom implementation of the algorithm provided in \cite{Yaylali} (more details are reported in the Methods section). Results are shown in Figure \ref{fig:Porkchop}, where the hyperbolic excess velocities at departure v$_\infty$(Earth) and arrival v$_\infty$(Mars/Venus) are plotted for a set of departure dates and times of flight (${\mathrm{TOF}}$). In general, the ${\mathrm{TOF}}$ obtained by using laser propulsion technology can be extremely reduced with respect to classical missions, namely not laser propulsion based. For instance, simulations reveal that a journey to Mars with departure on April 25$^{th}$ 2033 lasting 85 days would require a trusting phase of 69 s or 105 s for a tangential and a radial impulse respectively (solution labeled S$_{1}$ in Figure \ref{fig:Porkchop} and Table \ref{tab:Parameters}). If the journey lasts 120 days, the lightsail is propelled for 42 s or 75 s (solution S$_{2}$ in Figure \ref{fig:Porkchop} and Table \ref{tab:Parameters}), and if it lasts 200 days, the lightsail is propelled for 34 s or 65 s (solution S$_{3}$ in Table \ref{tab:Parameters}). Thus, the travel times are lower if compared with the typical ones obtained by Hohmann transfers, which are 259 days average in case of Mars. Similarly, to reach Venus by departing on January 25$^{th}$ 2033 in a ${\mathrm{TOF}}$ of only 50 and 100 days, the lightsail needs to be propelled for 103 s or 141 s in the first case, and 43 or 76 s in the latter (see solutions S$_{4}$ and S$_{5}$ in Figure \ref{fig:Porkchop} and Table \ref{tab:Parameters}). For comparison, it is worth to note that the average time of a Venus Hohmann transfer settles around 149 days. These results show also that the ligthsail need to be illuminated only for tens of seconds, so that the thruster time lasts only for a very limited time with respect to the journey; the lightsail can thus be released from the paylod after the propulsion phase, relieving also the mass of the nanosatellite. In Figure \ref{fig:Earth2planet} the simulation of spacecraft transfer orbits to Mars (left) and Venus (right) for the scenarios reported in Table \ref{tab:Parameters} are shown. The simulations are performed by making a propagation of the position of the bodies (Earth, Mars/Venus and nanosatellite) in the gravitational field of the Sun using a custom implementation of the algorithms presented in \cite{Curtis2020} (see Methods section).
In \cite{Sheerin2021} ${\mathrm{LEP}}$ is proposed as a potential technology capable to propel a cubesat in the Solar System, reaching Mars in a ${\mathrm{TOF}}$ comparable with ${\mathrm{DELP}}$. However, this technology is not very suitable for launching swarm of nanosatellites as each spacecraft needs to be equipped with its own proper motor, hence determining a mission cost increase. Instead, when applied to a single nanosatellite, ${\mathrm{LEP}}$ requires a laser source carrying a power orders of magnitude lower than in the case of ${\mathrm{DELP}}$ technology, which makes this technology competitive. This is due to the higher efficiency in converting the laser-transferred energy into thrust. However, there are few aspects that need to be carefully addressed. When ${\mathrm{LEP}}$ propulsion is considered, a non-negligible percentage of the spacecraft mass (i.e., typically $> 60\%$ or more) is needed for the high-power electric motors and its fuel. Furthermore, in order to transfer all the energy necessary to obtain the desired $\Delta V$, the spacecraft has to be continuously illuminated by the laser for days (e.g., 5 days for a 4 kg payload and a total $\Delta V=5$ km s$^{-1}$), making the spacecraft pointing system a rather challenging task (during the illumination phase, the spacecraft moves of tens of thousands of km while Earth is rotating). Vice versa, in case of ${\mathrm{DELP}}$, all the required thrust is given from the photons momentum transfer process, without the need of propellant or motors. Furthermore, by appropriately designing the laser array and sail sizes, the target $\Delta V$ can be achieved with a laser illumination of tens of seconds or, at most, of few minutes. During this time, the spacecraft typically moves of hundreds of km. This is certainly an aspect to take into account as it certainly simplifies the spacecraft pointing system. Reasoning in terms of total energy spent to keep the laser active and considering the mission scenarios reported in  \cite{Sheerin2021}, both technologies would spend hundreds of GJ kg$^{-1}$, with with a small saving in the case of ${\mathrm{LEP}}$ with respect to ${\mathrm{DELP}}$ when the target $\Delta V$ value is lower than $\simeq 15$ km s$^{-1}$. Finally, ${\mathrm{DELP}}$ presents lower mission risks, as the loss of a satellite is less likely as the pointing system is less complex. Even assuming the loss of a satellite as a possible scenario, the cost of ${\mathrm{DELP}}$-driven units is lower than for ${\mathrm{LEP}}$-driven ones. Further discussion is needed to compare both technologies in case the mission profile foresees the launch of a single massive satellite towards a planet.

\begin{table}
    \centering
    \begin{tabular}{l c|c c c|c c}
    \toprule
         \textbf{Parameter} & \textbf{Units} & \multicolumn{3}{c}{\textbf{Mars}} & \multicolumn{2}{c}{\textbf{Venus}} \\
    \hline 
    Solution label & & S$_1$ & S$_2$ & S$_3$ & S$_4$ & S$_5$ \\
    Departure date &  &\multicolumn{3}{c|}{April 25 2033} & \multicolumn{2}{c}{January 25 2033} \\
    Time of Flight & days & 85 & 120 & 200 & 50 & 100  \\ 
    v$_\infty$ at departure & km s$^{-1}$ & 6.30 & 4.01 & 3.12 & 8.90 & 4.03  \\
    Laser thrusting $\lVert$ & s    & 69   & 42   & 34   & 103  & 43    \\
    Laser thrusting $\bot$ & s    & 105  & 75   & 65   & 141  & 76 \\
    v$_\infty$ at arrival & km s$^{-1}$ & 12.50 & 6.96 & 3.33 & 16.58 & 4.01  \\
    \toprule
    \end{tabular}
    \caption{Parameters for an orbital transfer from Earth to Mars and to Venus in a 2033 launch window.  }
    \label{tab:Parameters}
\end{table}

Another important aspect to consider is the velocity at arrival to be selected according to the mission profile, as it could result in  a flyby, or in the insertion in the planetary orbit by means of proper maneuvers \cite{Duplay2022}. At the end of the transfer part, depending on the strategy adopted during approach, the spacecraft could be ballistically captured in a highly irregular orbit, which requires at least an high thrust maneuver to stabilize the orbit itself and to reduce the eccentricity. 
Preliminary calculation on insertion maneuvers consider $v_\infty=0$ with respect to the target planet at the arrival. The velocity budget has been estimated using GMAT suite \cite{GMAT}) to be $\Delta v \simeq 900 - 1400$ m s$^{-1}$, depending on the desired final orbit eccentricity and altitude. A chemical thruster with about 3 N thrust would allow to perform a sufficiently fast maneuver. In this scenario, the mass of the nanosatellite is estimated to be increased by a wet mass of 5 kg; moreover, an increase of the mass of reaction wheels needs to be taken into account given the total mass increment.

\section{Payload and on-board systems}

The perspective of launching a swarm of nanosatellites for heliosphere and planetary investigation with cost-effective laser propulsion requires a dramatic reduction of the payload mass and, in general, of the overall complexity of the spacecraft. The capability to send a high number of low cost satellites has the key advantage of redundancy, and the risk that some units experience a failure is largely compensated by the advantage of having multiple copies. When possible, commercially available \textit{Components-Off-The-Shelf} (${\mathrm{COTS}}$) have to be preferred since they have been proven to achieve high performance at increased functional density and low cost. However, they must be selected considering their radiation hardness, and, when necessary, tests should be carried out to prove their reliability \cite{sweeting2018}; in particular, semiconductor electronic components and integrated circuit devices (such as memories and microprocessors) should be tested in term of Total Ionization Dose (${\mathrm{TID}}$) and Single Event Effects (${\mathrm{SEEs}}$). The radiation hardening techniques involve logical (error-correcting code) and physical (shielding or redundancy) approaches; the latter one will increase the weight of the payload.

The selection and characteristics of subsystems are addressed according to the specific mission requirements, which is outside the scope of the present work. However, a first effort to identify the major subsystems of a payload for an interplanetary scientific mission is carried out \cite{ridenoure2016testingTL,burkhardt2018project}. First of all, the scientific nature of space exploration requires the adoption of a measuring system. As an example, here we refer to a specific sensor, being a planar Langmuir ion probe, which could be used to map the heliosphere plasma or to finely analyze a planet close environment. These type of sensors are very compact, as they consist of a fixed-bias flat-plate probe coupled with a dedicated electronics for the measurement of the collected ion current \cite{Bishop2017TheLI}. The weight is approximately 50 g for the envelope, electronics and sensor, while in terms of power the absorption peak is of 250 mW (Table \ref{tab:massbudget}). 

A communication system able to acquire and manage the data provided by the sensor and to transmit them to Earth must be carefully selected. In particular, telecommunications at deep space distances (i.e. $>0.5$ au) require the use of a high gain antenna (${\mathrm{HGA}}$) operating in either X-band or Ka-band \cite{Albugasem2021}, with reflectarray (${\mathrm{RA}}$) and mashed reflector (${\mathrm{MR}}$) antennas being the only solutions in order to combine the ability to support any polarization, high efficiency, and low mass (i.e., an aerial mass density $\simeq 1-1.5$ kg m$^{-2}$) \cite{Davarian2020}. ${\mathrm{RA}}$ antennas are characterized by low cost and very high reliability, but they are characterized by small bandwidths and typically support single-frequency links. On the contrary, MR antennas have large bandwidth allowing multiple-frequency operations, but have higher costs and are more prone to reliability issues. For instance, a MR antenna of only $D_\mathrm{t}=30$ cm in diameter provides a gain $G_\mathrm{t}$ higher than $25$ dBi in X-band and $34$ dBi in the Ka-band. Assuming a reasonable aerial mass density of 1.5 kg m$^{-2}$, the mass of this antenna can be estimated to be $\simeq 100$ g. Additional 100 g come from the mass of communication system electronics and cables, bringing the total weight of the telecommunications system to 200 g (Table \ref{tab:massbudget}). In order to evaluate a potential data transmission rate, the case of a satellite transmitting power of $P_0=$ 2 W and of a 70-m ${\mathrm{NASA}}$ Deep Space Network (${\mathrm{DSN}}$) receiver is considered. Such antenna gain $G_\mathrm{r}$ is $\simeq 74$ dBi in the X-band and $\simeq 83$ dBi in the Ka-band, so that the estimation of the received signal power $P_\mathrm{r}$ can be obtained by the Friis transmission equation:
\begin{equation}
    P_\mathrm{r}=P_\mathrm{0}G_\mathrm{t}G_\mathrm{r}\left( \frac{c}{4 \pi f d} \right)^2
\end{equation}
where $f$ is the carrier frequency and $d$ is the satellite-receiver distance. In the mission scenarios discussed in this work, a distance $\simeq 2$ au need to be considered, so that the power of the received signal with a 70-m ${\mathrm{DSN}}$ antenna is $\simeq -148$ dBm. The typical X-band noise temperature of such antenna is $T_\mathrm{0}=25$ K \cite{clauss2008}, which allows to achieve a Signal to Noise Ratio (${\mathrm{SNR}}$) >3 dB in down-link. With such ${\mathrm{SNR}}$ value, an appropriate choice of the modulation allows a down-link rate always higher than 3 kbps. 
The maximum antenna pointing error can be estimated by using the classical formula used for the diffraction from a circular aperture \cite{Wertz1999}. In particular, the $-3$ dB beam-width of a ${\mathrm{MR}}$ antenna with circular shape is $\theta_\mathrm{-3dB}\simeq 70 \frac{c}{fD_\mathrm{t}}=8.3^\circ$. If a maximum power loss of $L_\mathrm{\theta}=0.1$ dB is acceptable, the maximum tolerated pointing error $\theta_\mathrm{P}$ is estimated to be

\begin{equation}
\theta_\mathrm{P}\simeq \theta_\mathrm{-3dB}\sqrt{\frac{L_\mathrm{\theta}}{12}}=0.75^\circ
\end{equation}

To correct the attitude and pointing for communication purposes an \textit{Attitude Determination and Control System} (${\mathrm{ADCS}}$) can be used. It includes attitude sensors to determine the orientation, actuators to modify the attitude and reject disturbances, and a digital control link between the two. The vast majority of nanosatellites in orbit make use of simple magnetometers and Sun sensors to determine their pointing. Recently, high-performance micro-electromechanical systems (${\mathrm{MEMS}}$) detectors and actuators have started to be used for high precision pointing, such as digital sensors, star trackers and gyros \cite{Liddle_2020}; these new technologies offer the advantages of requiring limited volumes, weights and power consumption. The lack of a relevant magnetic field in the interplanetary space, and even in most planetary orbits, guides through the use of a reaction wheel-based attitude control systems in place of magnetic field actuators as magnetotorquer, with low-mass miniaturized reaction wheels already available in the market \cite{blueCanyon2022}. The commercial use for 1U cubesat, about the size of the proposed payload, assures that 3 micro reaction wheels with 3 mN s can be used, having a mass lower than 50 g per wheel. 
The use of reaction wheels requires momentum management in order to desaturate once the maximum speed is reached. The desaturation control actuator is normally obtained using magnetic torque coils, that are, as said, not applicable in lunar, martian or interplanetary orbit.
Few alternative desaturation strategies for a reaction wheel precise pointing are available. In particular, the use of thrusters should be considered. For example, small water resistojet, which can achieve the necessary desaturation total impulse, have been recently developed for nano-satellite application in order to allow small maneuvers, such as collision avoidance and attitude corrections. Commercially available technologies have a very compact size (cube of 20 mm side) and weight of $\simeq 40$ g for each thruster, including propellant. The number of thrusters necessary to perform the desaturation task has been estimated to be 6, considering redundancy. The use of thrusters for attitude control without the reaction wheel is a possibility in case of loose pointing requirements. In particular, the pointing accuracy required by the telecommunication antenna does not appears particularly stringent and it could be satisfied even without reaction wheels.\\
The \textit{Command and Data Handling} (${\mathrm{C\&DH}}$) system is responsible for supporting the ${\mathrm{ADCS}}$ and the collection and communication of the data acquired by the sensors to the Earth. Heritage from previous missions will serve as a basis for the selection of high performance and reliable microprocessors to execute the flight software, to manage the subsystems, and to simultaneously interface with the sensors \cite{freeman2020}.\\
Electrical power generation systems for nanosatellites have seen a continuous improvement from few watts in the past up to $10 - 20$ W available nowadays. Solar panels can be allocated along the sides of the spacecraft or can be deployed after the launch \cite{Liddle_2020,klesh_2013}. Currently, typical power densities for solar panels are in the $46-160$ W kg$^{-1}$ range at 1 au from the Sun \cite{Yost2021}. By re-scaling such values for the Mars distance, where the solar irradiance is about 65\% of that on Earth, the power densities to be considered are in the $30-104$ W kg$^{-1}$ range. Taking into account the mean power consumption estimated for all the satellite systems and summarized in Table \ref{tab:massbudget}, a conservative determination of the solar panels mass can be obtained by adopting the minimum value of power density (30 W kg$^{-1}$). As the total estimate of nanosatellite consumption is of about 3 W, most of which used by ${\mathrm{ADCS}}$ and communication systems, the correspondent solar panel mass is calculated to be 100 g. In addition, a battery is required for the management of peak consumption (e.g., desaturation and communications) and to support the mission during eventual solar power interruption when attitude maneuvers or Sun eclipses are in place. The design of the battery capacity requires the definition of the energy that needs to be stored and the maximum tolerated depth of discharge (${\mathrm{DoD}}$); reasonable values are two times the absorbed mean power for the capacity (i.e. $\simeq6$ Wh) and a ${\mathrm{DoD}}$ of 60\%. For instance, by using Li-ions batteries, which show an energy density value of 150 Wh kg$^{-1}$ with a inherent efficiency $\eta_\mathrm{bat}=0.95$ \cite{Yost2021, knap2020}, the total mass required by the battery is 100 g. \\
Table \ref{tab:massbudget} summarizes the weight and power estimates considering a list of technologies available on the market, mainly selected among ${\mathrm{COTS}}$; however, when needed, the reliability and radiation hardness may require particular space certified components. 
\begin{table*}[tpb]
    \centering
    \begin{tabular}{l l c c }
        \toprule
        \textbf{Subsystem}          & \textbf{Technology}    & \textbf{ Weight [g]}  & \textbf{Power [mW]}    \\
       
                \hline 
        Sensor                     & PLP                   & 50                    & 250  \\
        
                \hline 
        \multirow{2}{*}{Communications}         & Transmitter               & 100                     &  2000                     \\ 
                                                & Receiver                  & 100                                          \\ 
       
        \hline 
        \multirow{4}{*}{ADCS}       
                                            & Sun sensor (coarse)   & 40                    &  100                    \\
                                    & Sun sensor (fine)     & 5                     &  40                      \\
                                    & Reaction wheels (3)        & 150                   &  300                     \\
                                    & Desaturation thrusters   & 240                     &  30                      \\

        \hline 
        \multirow{2}{*}{C\&DH}      & Core PCB              & 5                     &  -                     \\ 
                                    & CPU                   & 1                     &  5                      \\

\hline 
        \multirow{3}{*}{T/S}      & alloy support              &  100   & \\
                                    & filler/harness & 20 &   \\
                                    & blanket           & 60 &  \\    
        \hline 
        
        \multirow{3}{*}{Power}      & Solar array           & 100                     &  (-3000)*                      \\ 
                                    & Battery               & 100                                           \\
                                    & PPU                   & 20                     &  50                      \\
         \hline 
       Total estimate             &                       & 1097                  & 2920  \\
        \toprule
       
        \end{tabular}
        
    \caption{Mass and power budget considering the mean power consumption. The power system is sized according to the estimated total and it is expressed as a negative number computed at the Mars distance*, not included in the total sum.}
     \label{tab:massbudget}
\end{table*}

In the case of the ${\mathrm{DELP}}$ technology, the lightsail needs also to be included in the mass budget. The weight of the lightsail depends on the membrane substrate, the nanostructure functionalizing it, and the boom supporting structure. In the present work, the thin film technology is proposed for the lightsail realization \cite{Santi2022}. Thin film of dielectrics are deposited on a substrate, which can be realized as a 10 $\mu$m thick Kapton membrane, or, even lighter, as a few hundreds of nm Silicon Nitride membrane. Many possible designs of boom structures exist, though the most used one has the main sail divided into four sections (or petals) and connected in a cross configuration. The boom can be realized in carbon-fiber-reinforced plastic \cite{trofimov2018}. For instance, in a worst-case scenario, a $S=10$ m$^2$ area sail made of such Kapton membrane (density of 1420 kg m$^{-3}$) weighs 140 g, and requires 8.94 m of booms with a weight of 134 g if the shape is square or 7.14 m of booms with a weight of 107 g if the shape is round; the multilayer coating contribution is neglected. In general, the total mass $m_\mathrm{l}$ of a lightsail, including substrate and booms, can be estimated by:

\begin{equation}
    m_\mathrm{l} = S\rho_\mathrm{s} t_\mathrm{s} + 2 \alpha \sqrt{S}\cdot\Tilde{\rho}_\mathrm{b}
\end{equation}
 
 where $\rho_\mathrm{s}$ and $t_\mathrm{s}$ are the density and thickness of the substrate membrane, respectively, and $\Tilde{\rho}_\mathrm{b}$ is the mass per unit length of the booms expressed in kg m$^{-1}$ i.e. 0.015 kg m$^{-1}$ in the present study). The parameter $\alpha$ depends on the shape of the lightsail: for a squared shape $\alpha=\sqrt{2}$ whereas for a round shape $\alpha=\sqrt{\frac{4}{\pi}}$. During the laser thrust phase, this weight needs to be taken into account for the estimation of the total mass, while it must not be accounted once the sail has been detached. In the case of ${\mathrm{LEP}}$ technology, the satellite needs to include the electric motor and an adequate photovoltaic system, which increase the total mass of 2.5 times \cite{Sheerin2021}.

\section{Lightsail}

\begin{figure*}
    \centering
   \includegraphics[width=1\textwidth]{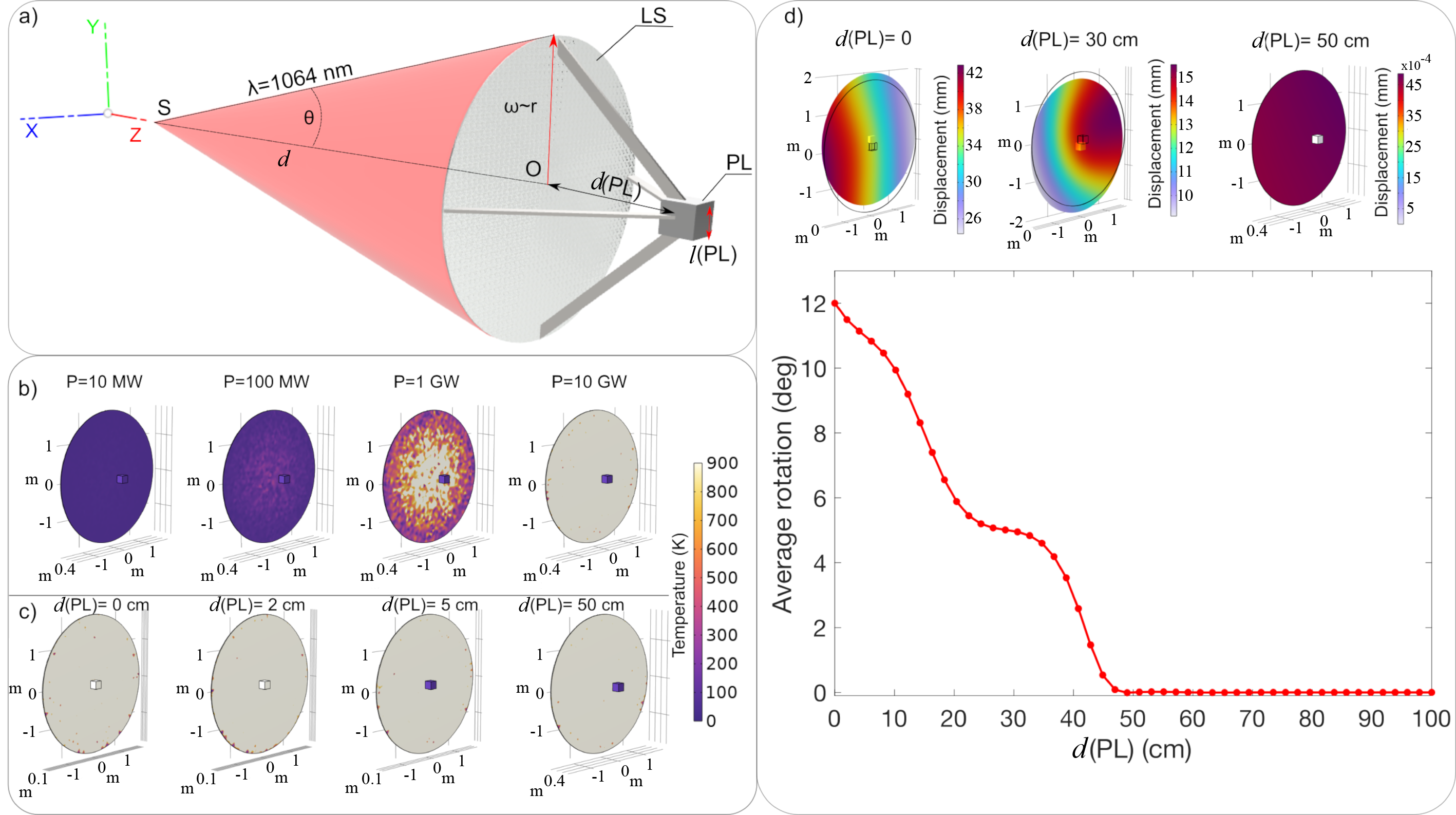}
    \caption{a) Schematic representation (not in scale) of the nanosatellite, comprising the lightsail ($\mathrm{LS}$) and the payload ($\mathrm{PL}$). b) Temperature analysis for different laser powers $P$, being 10 MW, 100 MW, 1 GW and 10 GW, in the case $d(\mathrm{PL})=50$ cm. c) Temperature analysis for $P=10$ GW varying the distance $d(\mathrm{PL})$. d) The maps (top panel) and the plot (bottom panel) report the deformations and the average rotation, respectively, occurring on the lightsail as function of the distance $d(\mathrm{PL})$.}
    \label{fig:therm_mech}
\end{figure*}

A basic requirement is that nanosatellites are thermally and mechanically stable during the acceleration phase realized by means of the high power laser beam. Some recent papers discuss the thermal and mechanical stability of thin layer(s) \cite{Santi2022} and nanostructured lightsails \cite{Taghavi2022}, but detailed analyses considering a configuration where the lightsail is part of a complete spacecraft comprising a small payload have not been reported so far. In the following discussion, a simple and efficient design is adopted for the ligthsail, consisting of a reflective thin film membrane (in particular TiO$_2$) \cite{Santi2022}. Even though nanostructured lightsails offer the advantage that sophisticated surface patterns could be optimized to guarantee high heat dissipation and mechanical stability, a thin film design allows to reduce the overall complexity and manufacturing costs, still ensuring a good efficiency in terms of propulsion and thermal management. Here, the previous analyses on thin film lightsails are extended considering a multiphysics approach (Finite Elements Methods - COMSOL Multiphysics - see Methods Section) where different aspects are investigated at the same time, i.e., thermal and mechanical behaviours taking into account the expected optical response. A sail made by TiO$_2$, having a radius of 1.8 m (i.e. a total area of 10 m$^2$) and a thickness of 1 $\mu$m is considered for simulation. As demonstrated in \cite{Santi2022}, the optical characteristics the TiO$_2$ thin film fully determine the thermal properties of the lightsail, which, in that case, was optimized for interstellar travel. In order to perform the present simulations, the optical constants used are those reported in \cite{Santi2022}, and in particular the extinction coefficient $k$ is assumed to be of $1\cdot10^{-6}$. In addition, for the finite element method simulations, the following material parameters have been used: an expansion coefficient of $10.2\cdot10^{-6}$ K$^{-1}$, Young's module of 150 GPa, Poisson's ratio equal to 0.26, and heat molar capacity equal to 58.2 J$\cdot$mol$^{-1}\cdot$K$^{-1}$.
Jointed to the sail, a Si cube of $l(\mathrm{PL})= 20$ cm side and 1 kg weight is considered as a representative payload (PL), as sketched in Figure \ref{fig:therm_mech}a.  In the simulations, the lightsail and the payload are considered rigidly connected at distances $d(\mathrm{PL})$ ranging from 0 up to 100 cm. A high-power laser of $\lambda_0$ = 1064 nm is used as source to push the sail. During the trusting phase, the sail should be ideally uniformly illuminated; however, here we consider a worse case scenario, in which the power distribution on the lightsail has a Gaussian profile, with a radius equal to the sail one. Indeed, the beam distribution at various distances will depend on the nature of the source and technology used to create a wide laser beam, but in the present simulations the illumination have been practically obtained using a very simple model, a single mode laser with divergence of 7.5$\times10^{-8}$ rad and a lightsail at $d = 50.000$ km. Different laser powers produce not only different pressure on the sail, but also different thermal effects, which can affect the operations and stability of the nanosatellite. In order to investigate this issue, different laser powers have been considered, being 10 MW, 100 MW, 1 GW and 10 GW. This last case corresponds to an irradiance of $I_0=1$ GW m$^{-2}$ on a 10 m$^2$ lightsail, which could be assimilated to the case discussed in section 2. In Figure \ref{fig:therm_mech}b the thermal heating of the lightsail is reported as function of the laser power for the case d(PL)=50 cm. For powers of 10 MW and 100 MW, the lightsail has a temperature close to $T \simeq$ 50 K, while for a laser power of 1 GW the temperature reaches an average less than $T \simeq$ 600 K. Finally, for a laser power of 10 GW the lightsail is uniformly heated, reaching a value of $T \simeq$ 900 K, still sustainable by the material \cite{Santi2022}.  %%

The significant heating produced by the high power laser beam is not surprising and can have a not negligible impact on the payload. Therefore, simulations have been performed at different distances $d(\mathrm{PL})$ in the case of $P=10$ GW. The results show that the silicon payload reaches temperature $T\simeq$ 900 K for $d(\mathrm{PL})<5$ cm, as illustrated in Figure \ref{fig:therm_mech}c, while for $d(\mathrm{PL})>50$ cm the payload can be considered thermally isolated from the sail. The thermal effects on the $\mathrm{PL}$ can also induce  displacement and rotations, hence instability, during the lightsail thrusting; hereafter, the term displacement indicates all deformations externally induced on the lightsail \cite{Taghavi2022}. This physical quantity and related rotation effects are numerically simulated. The results show that for $d(\mathrm{PL})$ up to 30 cm the entire systems has a non negligible displacement; on the contrary, for a $d(\mathrm{PL})=50$ cm, the entire body of the sail moves without any perturbation (displacement $\sim 4\times10^{-3}$ mm\,) (Figure \ref{fig:therm_mech}d, top panel). It is thus interesting to evaluate the dependence of the rotation from the distance between the payload and the lightsail. Numerical study confirms that for a $d(\mathrm{PL})>50$ cm the lightsail is not affected by any spin (Figure \ref{fig:therm_mech}d, bottom panel). As said, this analysis represents a worst case scenario, as the use of a larger laser beam waist/size is expected to reduce these effects.

\begin{figure*}
    \centering
   \includegraphics[width=1\textwidth]{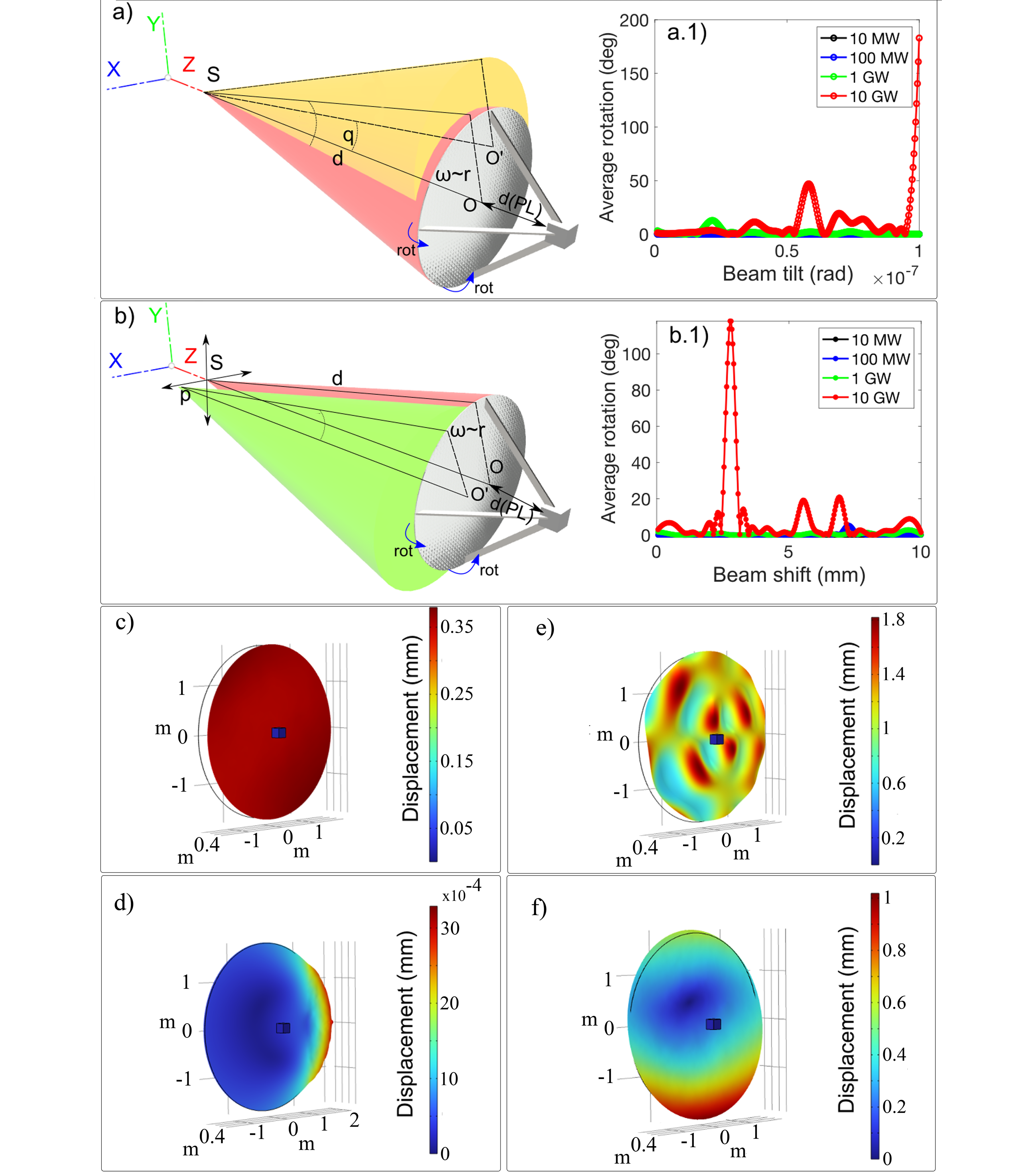}
    \caption{a) Representative sketches (not in scale) of the beam tilt $q$, indicated in yellow. a.1) Average rotation calculated by varying the laser power $P$, and $q$ from 0.1$\cdot 10^{-7}$ up to 1$\cdot 10^{-7}$ rad. b) Representative sketch (not in scale) of the source shift $p$ illustrated in green. b.1) Average rotation calculated by varying the laser power $P$ and for $p$ varying from 10 $\mu$m up to 10 mm. c) Displacement map for beam tilt 1$\cdot 10^{-7}$rad. d) Displacement map for $q=0.35\cdot 10^{-7}$ rad. e) Displacement map for $p=0.035$ mm. f) Displacement map for beam $p=3.2$ mm.}
    \label{fig:therm_diff}
\end{figure*}

\begin{figure*}
    \centering
   \includegraphics[width=0.9\textwidth]{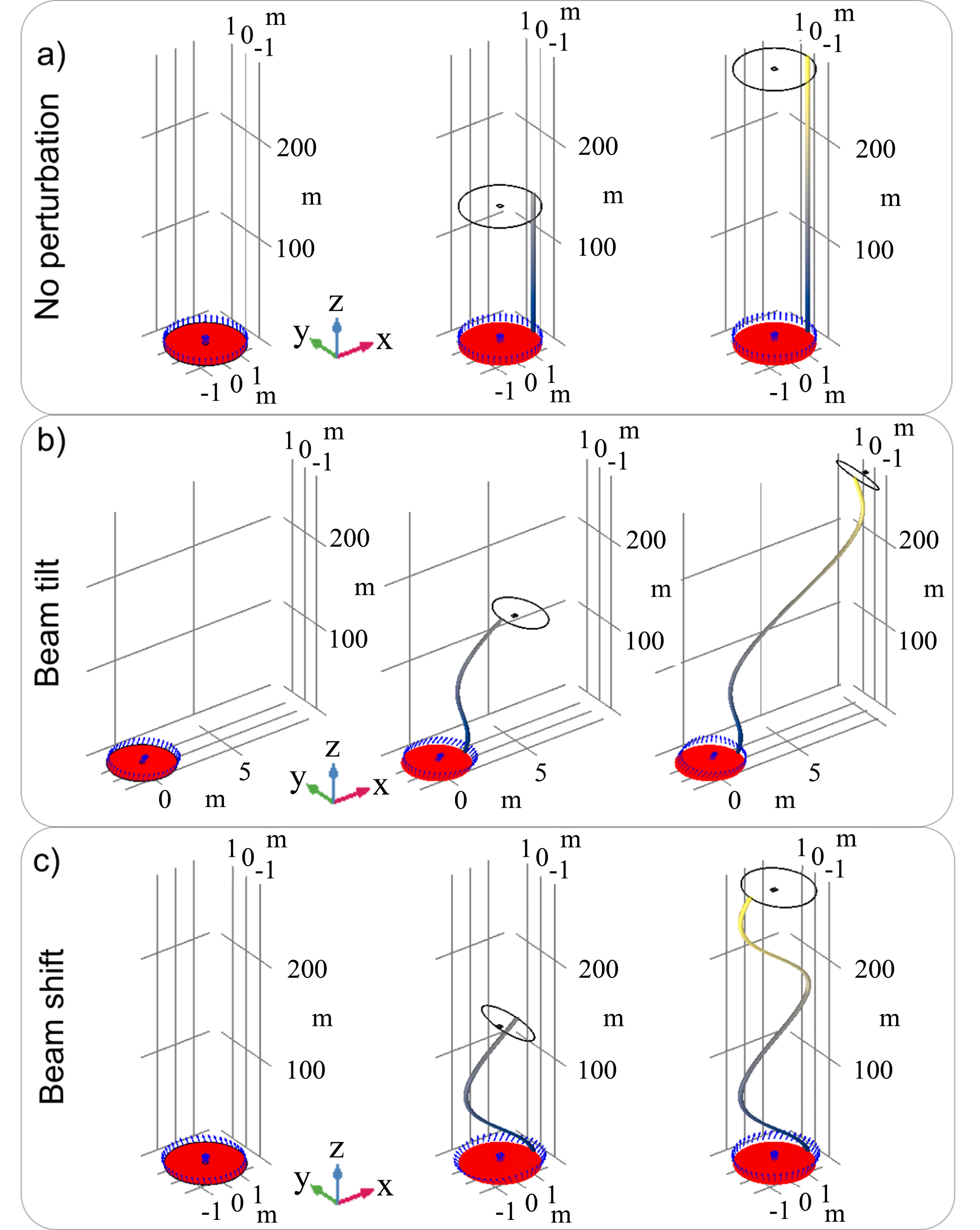}
    \caption{Numerical investigation of the lightsail trajectory; a) the laser beam is nominal; b) the laser beam is tilted of $q=1\cdot 10^{-7}$ rad; c) the beam is shifted of $p=2.8$ mm. For each result the initial lightsail position is indicated by the solid red object, while the black wireframe and blue arrows indicate the new position and the velocity direction along the new trajectory (gray-yellow tube gradient) respectively. For all the case studies $P=10$ GW. Each reported panel use the following axis scale factor: 2 along $X$ and $Y$, and 0.1 along $Z$.}
    \label{fig:therm_beam}
\end{figure*}

Another aspect to consider regards the misalignment between the beam Poynting vector and the lightsail normal. In detail, these numerical simulations are done considering a slightly beam tilt $q$ with respect to the thrust axis and a source shift $p$ in the plane perpendicular to the thrust axis, as depicted in \ref{fig:therm_diff}a and \ref{fig:therm_diff}b respectively. The lightsail average rotation for different laser beam powers and a beam tilt $q$ ranging from 0.1$\cdot 10^{-7}$  up to 1$\cdot 10^{-7}$ rad is reported in Figure \ref{fig:therm_diff}a.1. In detail, the numerical study highlights that for $P=10$ GW and $q=1\cdot 10^{-7}$ rad the average rotation is 180$^\circ$, while for the same power but $q=0.35\cdot 10^{-7}$ rad no rotation occurs. 
Moreover, the lightsail dynamic obtained by varying the laser power and by applying a source (S) shift $p$ from 10 $\mu$m to 10 mm has been evaluated (Figure \ref{fig:therm_diff}b). For instance, with $P=10$ GW and $p=2.8$ mm the lightsail is subjected to a rotation of 120$^\circ$ (Figure \ref{fig:therm_diff}b.1). In addition, the displacement analysis highlights that varying the beam tilt and shift the lightsail is subjected to several deformations, as illustrated in Figure \ref{fig:therm_diff}c-f.
The previous results led to a further interesting analysis, which shows how the lightsail trajectory changes consequentially to beam tilt $q$ or shift $p$.
In Figure \ref{fig:therm_beam}a, the simulation shows an unperturbed laser beam that impinges on the lightsail, and pushes it straight along its propagation direction $Z$. In the second case a laser beam tilted of $q=1\cdot10^{-7}$ rad produces a spinning of the lightsail and a drift into a new trajectory (Figure \ref{fig:therm_beam}b). Finally, in the last case the beam is shifted of $p = 2.8$ mm, inducing a lightsail rotation along the propagation direction (Figure \ref{fig:therm_beam}c). These numerical studies highlight how a laser misalignment produces considerable effects, which can results in spacecraft drifting and spinning, even to the point of sail crumpling. 
Other issues could be related to the atmospheric turbulence, which can play a major role in distorting the beam profile and thus the lightsail response. For the sake of simplicity this aspect will be disregarded here and will be considered in a more detailed study in future.
Overall, simulations demonstrate how complex is the realization and control of the illumination of the lightsail-payload system. Once the technology and design of the lightsail as well as the characteristics of the laser beam have been defined, it will be necessary to deeply study and simulate the various effects that can be induced by the radiation pressure, especially in terms of mechanical stability, and eventually develop an attitude control system to overcome potential detrimental effects. All this will be the subject of future studies.

\section{Discussion and potential applications}
The theoretical scenario of reaching relativistic velocities makes ${\mathrm{DELP}}$ technology extremely interesting for the exploration of deep space. However, tremendous technical challenges exist. In particular, the fabrication of a large area, extremely thin and stable lightsail appears out of reach due to current technology limitations. Furthermore, no clear strategies for slowing down a relativistic flying object exist, a crucial aspect when the object mission is performing observations or analyses \cite{Heller2017, Perakis2016_2}. Last but not least, the relativistic object needs to be equipped with a communication system of adequate size, mass and power to send data back to Earth. Therefore, seeking a feasible mission scenario, in this article we discuss how ${\mathrm{DELP}}$ can be applied for the exploration of the Solar System.
In particular, ${\mathrm{DELP}}$ technology is proposed for propelling swarms of nanosatellites into the Solar System, for example to map different areas of the heliosphere or a specific planetosphere. The effort of building a launch facility is justified by the multitude of nanosatellites that can be launched in sequence. With respect to deep space exploration, Solar System missions foreseen relaxed requirements even though they can be exploited for the development and test of key technologies, thus laying the groundwork for enabling the technologies for deep space exploration. In the present discussion, the ${\mathrm{DE}}$-${\mathrm{STAR}}$ and ${\mathrm{DELTA}}$ laser facility concepts have been exploited, where the laser sources are positioned either in orbit or on Earth surface, respectively. However, a third, groundbreaking perspective is to place the laser system on the Moon's surface, having great benefits such as the lower gravitational attraction with respect to the Earth, lack of atmosphere, no obstacles during the lighting phase and less risk of accidents. Although this solution is on a long term time scale, it is however very well in line with the desire to create a human colony on the Moon \cite{Creech2022}. 
Prior the launch and in case of orbital release of nanosatellites, a first phase of orbit insertion should be planned by exploiting commercially available deploying platforms which can be mounted on a large variety of launch vehicles and can be preconfigured to accommodate any kind of satellite from 1U up to assemblies of 1$\times$12U, 2$\times$6U and 4$\times$3U \cite{sweeting2018}. A sail deployment phase should also be foreseen according with previous studies \cite{Johnson2011, johnson2011_2, Fu2016, Vatankhahghadim2021}. Considering that the sail will be used to reach the targeted velocity within few hundreds of seconds and that no additional role for the sail is expected in the proposed mission, it could be reasonable to consider to detach the sail from the payload after the acceleration phase. This would enable an easier operation of the nanosatellites during the journey and at target. In any case, our preliminary simulations show that the first thrusting phase is very complex and delicate and that the modelling of all the details of the illumination system is fundamental to guarantee a successful departure and acceleration. In particular, once the illumination system technology is defined, and the laser beam properties given, a detail analysis on the mechanical stability of the system versus potential beam misalignment needs to be carried out. On the contrary, the thermal stability does not pose any concerns. 

The designs and simulations reported in this work show that the proposed technology can be a valid option to develop a family of small satellites, capable of operating also as a cooperative network, for large scale observations or for the mapping of the heliospheric environment. Indeed, in this last mission scenario the nanosatellites would be radially propelled without the need of further orbital maneuvers. To date, the interplanetary environment, and in particular the heliospheric plasma, is only partially known due to the few existing opportunities for carrying out in-situ measurements, basically linked to scientific exploration missions \cite{Plainaki2016}. The composition and characteristics of the heliospheric plasma remain defined mainly through theoretical models only partially verified. Therefore, there is an urgent need to perform a more detailed mapping of the heliospheric environment especially due to the growth of the human activities in space. Consequently, the potential use of a large set (swarm) and low cost nanosatellites equipped with optimized instrumentation can significantly improve the possibility of heliosphere characterization. 
Further exploitation can be the targeting of a planet such as Mars and to perform not just a flyby, but a capture. In this scenario, a swarm of nanosatellites can host different sensors, or it can be used for constituting a communication system network for a Martian colony. Finally, by increasing either the $S/m_\mathrm{T}$ ratio or the laser illumination time $t$ of a factor 5, values for $\Delta V$ much higher than $20 - 25$ km s$^{-1}$ can be obtained, enabling the possibility of reaching the helio-pause and even beyond in just few years (i.e. heliocentric distances of $\simeq 100 - 130$ au in less than 10 years). Although within these last mission scenarios there is the need to upgrade the communication system by adopting nuclear-type energy sources, resulting in higher mass budgets, higher costs and more restrictive safety protocols, the possibility of launching swarms of nanosatellites to explore the most remote part of the heliosphere would still remain attractive and an exciting frontier to reach.

\begin{acknowledgments}
PML funding for this program comes from NASA grants NIAC Phase I DEEP-IN – 2015 NNX15AL91G and NASA NIAC Phase II DEIS – 2016 NNX16AL32G and the NASA California Space Grant NASA NNX10AT93H as well as a generous gift from the Emmett and Gladys W. Fund for the Starlight program and from the Breakthrough Initiatives Foundation for the Starshot program.
\end{acknowledgments}

\section*{Author contributions}
G.S. developed the mathematical code and carried out all the astrodynamics simulations; he drafted some parts of the paper.
A.J.C. contributed to the analysis of various technical aspects, such as telecommunication and electronics, supervised the astrodynamics simulations, contributed to the writing and paper and to its finalization. He prepared the TOC figure. 
G.E.L. and D.G. conceived the thermal and mechanical analysis, contributed to relative simulations and drafted some parts of the paper.
M.M. designed the nanosatellite systems and subsystems.
P.L. provided insightful ideas and inputs, as well as valuable feedback.
G.F., M.B., G.P., N.A., L.S., D.P., L.B., R.P.Z., R.R contributed with fruitful scientific discussions. 
M.G.P. conceived the paper, coordinated the scientific activities, wrote the paper and finalized it.
All authors reviewed the final manuscript.

\appendix

\section{Methods}
\section*{Orbit simulations}
The astrodynamics solutions for Mars' and Venus' transfer orbits are computed by solving the Lambert's problem to determine the fastest trajectories with lower costs requirements in terms of departure and arrival velocities. The mission scenarios selected are computed for the year 2033, when both a perihelic opposition of Mars and an inferior conjunction of Venus occur; however, this choice is not binding and, in general, it is possible to envision equivalent mission scenarios in other years. The departure windows considered start on April 1$^{st}$ 2033 for Mars and January 1$^{st}$ 2033 for Venus; then, for the 60 following days, the transfer orbit solution is computed, obtaining the required $\Delta V$, which is directly linked to the laser illumination time by using Eq. \ref{eq:DV}, the hyperbolic excess velocity at departure and at arrival, as well as the ${\mathrm{TOF}}$ associated. The sets of solutions, obtained with a custom implementation of the algorithm provided in \cite{Yaylali}, are shown in the porkchop plots of Figure \ref{fig:Porkchop}. For the selected solutions, the spacecraft  and the planets trajectories are then simulated by numerically solving the differential equations of a 3D 4-body problem comprising the Sun-Earth-Mars/Venus-spacecraft system, the latter being massless. The numerical integration, performed at a step of 1 s, is based on the algorithm presented in \cite{Curtis2020}, which makes use of the \textit{ODE45} solver in ${\mathrm{MATLAB}}$. The evolution of the planets' and spacecraft's positions are computed in the \textit{J2000} reference frame, while the final orbits are projected in the \textit{J2000 ecliptic} reference frame for a better graphical visualization (Figure \ref{fig:Earth2planet}). The ephemeris of the planets at the departure time are computed by making use of the ${\mathrm{MATLAB}}$ Interface to the ${\mathrm{SPICE}}$ toolkit \cite{Mice2022} at the ${\mathrm{DE440}}$ integration epoch.

\section*{System mechanical and thermal stability simulations}
The COMSOL based model uses the "Geometrical Optics" interface to trace the paths of rays through the lightsail system. A Gaussian beam source is used and it is placed 50000 km far to the target. The "Heat Transfer in Solids" and "Solid Mechanics" interfaces are used to model the thermal expansion and the displacements/ deformations and rotations of the proposed system. In order to consider the rigid movement of the entire system lightsail and payload in the mechanics has been used a prescribed velocity of 10.000 km h$^{-1}$,  along Z and the mass center has been calculated and included in the numerical simulation. In order to determine any rotation that can affect the lightsail the "average rotation" node is added to the model. This node allows computing average rotation for a set of points (along the edges of the lightsail) with respect to the nanosatellite center of rotation.
The ray trajectories and temperature distribution affect each other through a bidirectional coupling. In other words, the ray trajectories affect the temperature field, which in turn perturbs the ray trajectories, both directly and through the resulting structural deformation. To solve for the ray trajectories and temperature in a self-consistent manner, the dedicated Ray Heating interface and Bidirectionally Coupled Ray Tracing study step are used. The Bidirectionally Coupled Ray Tracing study step sets up a solver loop in which the ray trajectories and temperature are computed in alternating steps for a fixed number of iterations. 
\nocite{*}

\bibliography{bibliography}% Produces the bibliography via BibTeX.

\end{document}